\documentclass[aps,prc,twocolumn,superscriptaddress,floatfix,nofootinbib,longbibliography]{revtex4-1}

\usepackage[utf8]{inputenc}
\usepackage[english]{babel}
\usepackage{amssymb,amsthm,amsmath,amstext,amsbsy,amsopn}
\usepackage{bbm}
\usepackage{graphicx}
\usepackage{isotope}
\usepackage{float}
\usepackage[dvipsnames]{xcolor}

\usepackage{hyperref}
\hypersetup{
    colorlinks = true,
    citecolor  = blue,
    linkcolor  = blue,
    urlcolor  = blue,
}

\newcommand{\la}{\langle}
\newcommand{\ra}{\rangle}

\newcommand{\gcrit}{g_\text{crit}}

\newcommand{\ie}{i.e.}
\newcommand{\eg}{e.g.}

\newcommand{\ecpt}[1]{\text{EC-PT}(#1)}

\newcommand{\Strain}{\ensuremath{D_\text{train}}}
\newcommand{\Snormal}{\ensuremath{\Strain^{<}}}
\newcommand{\Ssuperfluid}{\ensuremath{\Strain^{>}}}
\newcommand{\Smixed}{\ensuremath{\Strain^{\text{mixed}}}}

\newcommand{\ECnormal}{\ensuremath{\text{EC}(\Snormal)}}
\newcommand{\ECsuperfluid}{\ensuremath{\text{EC}(\Ssuperfluid)}}
\newcommand{\ECmixed}{\ensuremath{\text{EC}(\Smixed)}}

\newcommand{\Nec}{N_\text{EC}}
\newcommand{\Mec}{\mathcal{M}^\text{EC}}

\newcommand{\HF}{\ensuremath{|\Phi_\text{HF}\ra}}

\newcommand{\pCI}[1]{\ensuremath{\text{pCI-}#1\text{p}#1\text{h}}}

\begin{document}

\allowdisplaybreaks

\title{
Eigenvector continuation for the pairing Hamiltonian
}

\author{M.~Companys Franzke}
\email{mcompanys@theorie.ikp.physik.tu-darmstadt.de}
\affiliation{Technische Universit\"at Darmstadt, Department of Physics, 64289 Darmstadt, Germany}

\author{A.~Tichai}
\email{alexander.tichai@physik.tu-darmstadt.de} 
\affiliation{Technische Universit\"at Darmstadt, Department of Physics, 64289 Darmstadt, Germany}
\affiliation{ExtreMe Matter Institute EMMI, GSI Helmholtzzentrum f\"ur Schwerionenforschung GmbH, 64291 Darmstadt, Germany}
\affiliation{Max-Planck-Institut f\"ur Kernphysik, Saupfercheckweg 1, 69117 Heidelberg, Germany}

\author{K.~Hebeler}
\email{kai.hebeler@physik.tu-darmstadt.de}
\affiliation{Technische Universit\"at Darmstadt, Department of Physics, 64289 Darmstadt, Germany}
\affiliation{ExtreMe Matter Institute EMMI, GSI Helmholtzzentrum f\"ur Schwerionenforschung GmbH, 64291 Darmstadt, Germany}
\affiliation{Max-Planck-Institut f\"ur Kernphysik, Saupfercheckweg 1, 69117 Heidelberg, Germany}

\author{A.~Schwenk}
\email{schwenk@physik.tu-darmstadt.de}
\affiliation{Technische Universit\"at Darmstadt, Department of Physics, 64289 Darmstadt, Germany}
\affiliation{ExtreMe Matter Institute EMMI, GSI Helmholtzzentrum f\"ur Schwerionenforschung GmbH, 64291 Darmstadt, Germany}
\affiliation{Max-Planck-Institut f\"ur Kernphysik, Saupfercheckweg 1, 69117 Heidelberg, Germany}

\begin{abstract}
The development of emulators for the evaluation of many-body observables has gained increasing attention over the last years.
In particular the framework of eigenvector continuation (EC) has been identified as a powerful tool when the Hamiltonian admits for a parametric dependence.
By training the emulator on a set of training data the many-body solution for arbitrary parameter values can be robustly predicted in many cases.
Furthermore, it can be used to resum perturbative expansions that otherwise diverge.
In this work, we apply EC to the pairing Hamiltonian and show that EC-resummed perturbation theory is in qualitative agreement with the exact solution and that EC-based emulators robustly predict the ground-state energy once the training data are chosen appropriately.
In particular the phase transition from the normal to the superfluid regime is quantitatively predicted using a very low number of training points.
\end{abstract}


\maketitle

\section{Introduction}

In \textit{ab initio} nuclear structure calculations new challenges emerged over the last years.
The combination of basis-expansion methods with chiral effective field theory interactions has enabled routine computations of medium-mass nuclei~\cite{Herg20review,Hebe203NF}.
This has led to the quest to develop accurate interactions for medium-mass nuclei, \eg{}, by including medium-mass observables in the construction of new chiral interactions~\cite{Ekst15sat,Jian20N2LOGO}, and to explore uncertainty estimates from the effective field theory truncation and uncertainties in the low-energy couplings~\cite{Hu2021lead}.
However, both of these aspects require repeated solutions of the many-body problems when varying the underlying interactions.
To cope with these significant computational demands, emulators have emerged as powerful tools to efficiently lower the computational burden of performing explicit calculations by mimicking the true many-body solution.

For this purpose reduced basis methods (RBM) and more general model order reduction (MOR) techniques have been widely applied in the natural sciences to reduce the computational cost by projecting the problem onto an appropriately chosen subspace~\cite{Almroth1978,Parlett1998,Quarteroni2014,Quarteroni2015,Hesthaven2015,brunton_kutz_2019}. In nuclear theory, RBMs have been applied to density functional theory~\cite{Bonilla2022,Giuliani:2022yna}, to the solution of the Schr\"odinger equation for single-particle Hamiltonians~\cite{Anderson2022} and to scattering problems. For an overview over different MOR methods see Refs.~\cite{Melendez:2022kid,Drischler:2022ipa}. Eventually, MOR methods also enable uncertainty quantifications.
A particular variant of RBMs is the so-called eigenvector continuation (EC) approach. The EC framework has emerged as a particularly versatile method to emulate solutions of the many-body problem governed by parametric dependencies~\cite{Frame2018}.
Over the last years, the EC method has been applied to various few- and many-body problems, \eg{}, for uncertainty quantification in few-nucleon systems~\cite{Konig:2019adq} and medium-mass nuclei~\cite{Ekstroem2019}, for scattering and reactions~\cite{Furnstahl:2020abp,Melendez:2021lyq,Drischler2021,Zhang:2021jmi}, for finite-volume extrapolation~\cite{Yapa:2022nnv} and nuclear matter computations~\cite{Jiang:2022oba}. Moreover, EC has been used as a resummation tool for perturbative expansions enabling a robust extraction of many-body observables when the perturbation series diverges~\cite{Demol2020EC,Demol20BMBPT,Companys2021}.

Practically, the emulator construction is based on the solution of a generalized eigenvalue problem on a small subspace spanned by a set of non-orthogonal many-body states defining the training vectors. 
From that training manifold, high-dimensional parameter spaces can be exhausted at a tractable computational cost. 
While in few-body applications the training vectors are virtually exact, the development for many-body calculations is accompanied by the introduction of approximation schemes that facilitate the emulator construction.
While various applications from exact training vectors exist, the interplay of emulator quality and truncated training vectors has not been studied in detail.

This work is dedicated to a detailed study of the exactly solvable pairing Hamiltonian~\cite{Richardson1963,Richardson1964} that serves as a model for nuclear superfluidity and is a common test ground for novel many-body frameworks~\cite{HjorthJensen2017}.
We will first establish EC as a robust resummation tool for many-body perturbation theory (MBPT) even though the bare perturbative expansion breaks down. In the second part of this work, we investigate the accuracy of an EC-based emulator and its sensitivity to the selected training points\footnote{Note that in the MOR literature, training points are often referred to as snapshots.} used in the construction.
Recently, the same system has been studied using RBMs built from approximate density matrix renormalization group (DMRG) wave functions~\cite{Baran2022rbm}.

\section{Eigenvector continuation}
\label{sec:ec}

Eigenvector continuation provides a powerful framework for the construction of emulators whenever the Hamiltonian admits for a parametric dependence $H(g_1,...,g_N)$ with all $g \in \mathbb{R}$~\cite{Frame2018}.
By training the EC emulator on a small set of training points in the parameter space, it may robustly predict many-body observables for different parameter values without the need of performing explicit many-body simulations.
In nuclear structure applications such a tool is particularly powerful, because the number of low-energy couplings in chiral effective field theory interactions may be systematically varied with an accurate emulator~\cite{Ekstroem2019,Jiang:2022oba}.

The construction of the EC emulator is based on the definition of a set of $\Nec$ training points $\Strain = \{g_i\}$ that defines a manifold of training vectors\footnote{In our applications there is only a dependence on a single coupling. However, the discussion straightforwardly generalizes to the case of several couplings.}
\begin{align}
    \Mec (\Strain) = \{ | \Psi(g_i) \ra \, : \, g_i \in \Strain \} \, ,
    \label{eq:ECman}
\end{align}
where $|\Psi \ra$ denotes the many-body state of interest, \eg{}, the ground state.
Energies in the EC framework are evaluated at the target coupling $g_\circ \in \mathbb{R}$ by solving the generalized eigenvalue problem
\begin{align}
    H \vec{x} = E N \vec{x} \, ,
    \label{eq:geneig}
\end{align}
where the Hamiltonian and norm kernel are obtained for the training vectors as
\begin{subequations}
\begin{align}
    H_{ij} &= \la \Psi(g_i) | H(g_\circ) | \Psi(g_j) \ra \, , \\
    N_{ij} &= \la \Psi(g_i) | \Psi(g_j) \ra \, .
\end{align}
\end{subequations}
Since the number of (non-orthogonal) training vectors $\{|\Psi(g_i)\ra\}$ is much smaller than the size of the configuration basis in diagonalization approaches, the EC diagonalization is of very limited cost.
The solution of the generalized eigenvalue problem admits for $\Nec$ solutions, so that the EC approach provides also access to excited states belonging to the same symmetry class, \eg{}, spin or parity.
This was recently studied for the case of the anharmonic oscillator~\cite{Companys2021}.
Finally, we note that the EC approach can be used for any other operator $O$ instead of $H$, thus enabling calculations of other observables~\cite{Konig:2019adq,Ekstroem2019}. 

\section{Pairing Hamiltonian}
\label{sec:ham}

In this work we study the pairing Hamiltonian
\begin{align}
    H(g) = 
    \sum_{p}^\Omega \epsilon_p 
    (c^\dagger_p c_{p} + c^\dagger_{\bar p} c_{\bar p} )
    -
    g \sum_{pq}^\Omega c^\dagger_p c^\dagger_{\bar p} c_{\bar q} c_{q} \, ,
\end{align}
where $\bar p$ denotes the time-reversed state of $p$, $\Omega$ denotes the number of two-fold degenerate levels with single-particle energies $\epsilon_p$, and $g$ the (real-valued) strength of the two-body interaction.
The single-particle spectrum is taken to be equidistantly spaced $\epsilon_p = p \, \Delta \epsilon$, with a level spacing of $\Delta \epsilon = 1$ (in natural units).
The pairing Hamiltonian has an exact Richardson solution, which is accessible for arbitrary coupling values and system size~\cite{Richardson1963,Richardson1964}. The Richardson solution is obtained by solving a set of non-linear coupled equations for the pair energies $E_\alpha$, where $\alpha=1, \ldots, N_\text{occ}$ with the number of occupied pair states $N_\text{occ}$,
\begin{align}
    1-g\sum_{k=1}^{2\Omega}\frac{1}{2\epsilon_k-E_\alpha}-2g\sum_{\beta \neq \alpha=1}^{N_\text{occ}}\frac{1}{E_\alpha-E_\beta}&=0\, .
    \label{eq:rich}
\end{align}
Here, the first sum runs over all possible single-particle states, while the second sum is restricted to occupied pair states.
In the following, we will study the system at half filling, \ie{}, half of the shells are doubly occupied yielding a pair number of $N_\text{pair} = \Omega/2$.
The final ground-state energy is given by $\sum_\alpha E_\alpha$. 
In our numerical benchmark we employ the formalism described in Ref.~\cite{Claeys2015} to solve Eq.~\eqref{eq:rich}.

It is well known that the pairing Hamiltonian undergoes a transition from a normal to a superfluid phase once a critical (positive) coupling value $\gcrit$ is exceeded.
For $g> \gcrit$ the system is no longer well approximated by low-rank particle-hole excitations but dominated by the formation of Cooper pairs resembling the superfluid character of the system.
This transition can be qualitatively captured by using Bardeen-Cooper-Schrieffer (BCS) mean-field theory, where $U(1)$ particle-number symmetry is explicitly broken at the mean-field level to account for the pairing correlations in the system~\cite{Coop56pairs,RingSchuck80}.
Therefore, the corresponding BCS vacuum $|\Phi_\text{BCS} \ra$ is not an eigenstate of the particle-number operator $A$, but only restricted to have correct average particle number, \ie{}, $N_\text{occ} = \la \Phi_\text{BCS} | A| \Phi_\text{BCS} \ra$.
The emergence of pairing correlations is further accompanied by a breakdown of standard symmetry-conserving correlation expansions and requires the use of quasiparticle reformulations to account for the collective effects induced by the static correlations~\cite{Soma13GGF2N,Sign14BogCC,Tichai18BMBPT}.

In our benchmarks we employ Rayleigh-Schr\"odinger perturbation theory
using the partitioning
\begin{align}
    H(\lambda) = H_0 + \lambda H_1 \, ,
\end{align}
where $H_0$ is taken to be the (diagonal) one-body part of the pairing Hamiltonian.
The ground-state wave function is expanded in the series
\begin{align}
    |\Psi \ra =  \sum_{p=0}^\infty \lambda^p \, | \Phi^{(p)}\ra \, ,
\end{align}
where $|\Phi^{(p)}\ra$ denotes the $p$-th order state correction~\cite{Shav09MBmethod}.
Corrections to the ground-state energy are calculated via $E^{(p)} = \la \Phi^{(0)} | H_1 |\Phi^{(p-1)} \ra$. The reference state $| \Phi^{(0)}\ra$ is given by the Hartree-Fock (HF) state \HF{}~\cite{Tich16HFMBPT,Tichai2020review}.

Moreover, we employ the $n$-particle-$n$-hole pair configuration interaction (pCI-$n$p$n$h) framework, where the $A$-body configuration space is spanned by all pair excitations up to $n$-particles-$n$-holes from of the HF state \HF{}.
The low-lying spectrum is obtained from diagonalizing the many-body Hamiltonian using Lanczos techniques.
While this does not provide a scalable solution for large systems, it is numerically tractable for $\Omega \lesssim 20$.
In the case the maximum number of $n$-particle-$n$-hole excitations is given by $n=2N_\text{pair}$, the exact result (full CI, FCI) is recovered.

\section{Eigenvector continuation as a resummation tool}

Eigenvector continuation can be used to resum the perturbation series by performing the EC in the space of MBPT state corrections ~\cite{Demol2020EC,Demol20BMBPT}.
The Hamiltonian and norm matrices that enter Eq.~\eqref{eq:geneig} are then given by
\begin{subequations}
\begin{align}
    H_{pq} &= 
    \la \Phi^{(p)} | H | \Phi^{(q)} \ra \, , \label{eq:ecptH} \\
    N_{pq} &= 
    \la \Phi^{(p)} | \Phi^{(q)} \ra \, .
    \label{eq:ecptN}
\end{align}
\end{subequations}
where $N_{p0} = \delta_{p0}$ by intermediate normalization.
In this work, MBPT state corrections up to first order are included; therefore we denote this approach as \ecpt{1}.
Explicit expressions for the matrix elements in Eqs.~\eqref{eq:ecptH} and~\eqref{eq:ecptN} are evaluated by virtue of Wick's theorem~\cite{Wick50theorem}.
The simplest expression involves the second-order energy correction
\begin{align}
H_{10} = E^{(2)} = \frac{1}{8}\sum_{ai} \frac{g^2}{\epsilon^a_i}\, ,
\end{align}
where $\epsilon^a_i = f_i - f_a$ is a shorthand notation for the energy denominators (with $i,j$ referring to occupied hole states and $a,b$ to particle states) and $f_p =\epsilon_p - g n_p$ denotes the normal-ordered one-body part. The overlap is given by
\begin{align}
N_{11} = \frac{1}{16} \sum_{ai} \frac{g^2}{(\epsilon^a_i)^2} \, .
\end{align}
The most complicated expression corresponds to the matrix element between two first-order state corrections and consists of three contributions
\begin{subequations}
\begin{align}
H^{[0B]}_{11} &= E_\text{HF} \, N_{11} \, ,\\
H^{[1B]}_{11} &= 
-\frac{1}{8}\sum_{ai} \frac{g^2}{\epsilon_i^a} \, , \\
H^{[2B]}_{11} &= 
-\frac{1}{32}\sum_{ai} \frac{g^3}{\epsilon_i^a} 
\biggl(
\sum_b \frac{1}{\epsilon_i^b} +
\sum_j \frac{1}{\epsilon_j^a}
\biggr) \, , 
\end{align}
\end{subequations}
where the superscript indicates from which 0-, 1-, or 2-body part of the Hamiltonian the contribution originates. For the next order EC-PT(2), second-order state corrections that contain 4p4h-excited components are needed.

\begin{figure}[t]
    \centering
    \includegraphics[width=\columnwidth,clip=]{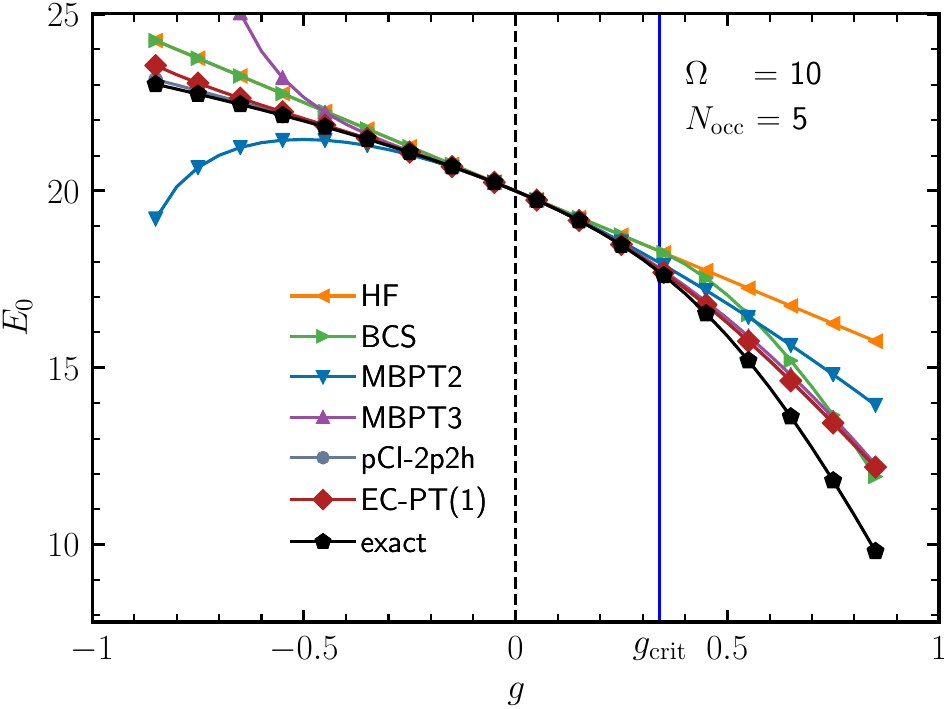}
    \caption{Comparison of different basis-expansion methods (for details see text) for the ground-state energy of the pairing Hamiltonian as a function of the coupling $g$ for $\Omega=10$ at half-filling $N_\text{occ}=5$. The critical coupling $\gcrit=0.34$ is indicated by the vertical line.}
    \label{fig:compare}
\end{figure}

\begin{figure}[t!]
    \centering
    \includegraphics[width=\columnwidth,clip=]{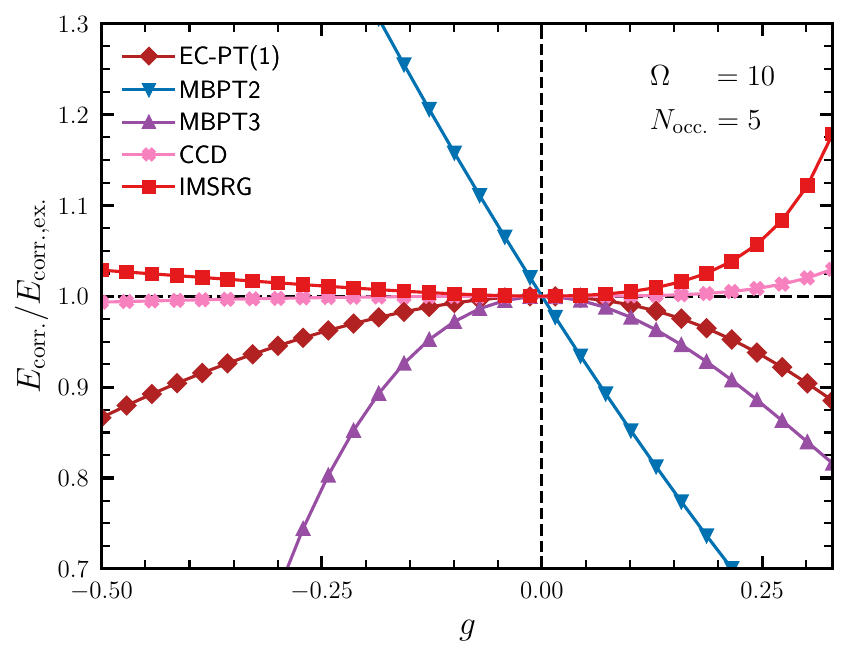}
    \caption{Relative error of the correlation energy compared to the exact result of basis-expansion methods, including coupled-cluster doubles (CCD) and in-medium similarity renormalization group (IMSRG) calculations, in the normal regime ($g<\gcrit$) for $\Omega=10$ at half-filling.}
    \label{fig:mbbenchmark}
\end{figure}

Figure~\ref{fig:compare} shows a comparison of different many-body methods for the solution of the pairing Hamiltonian. 
In the normal phase ($g<\gcrit$) the HF and BCS solution coincide, since the coupling strength is too weak to support a superfluid ground state~\cite{RingSchuck80,Duguet2017}.
For $g>\gcrit$ the BCS state explores the enlarged variational space, yielding a lower energy than the symmetry-conserving HF solution.
Both HF/BCS mean-field approaches lack all dynamical particle-hole correlations leading to the deviations from the exact solution with increasing coupling.

\begin{figure}[t]
    \centering
    \includegraphics[width=\columnwidth,clip=]{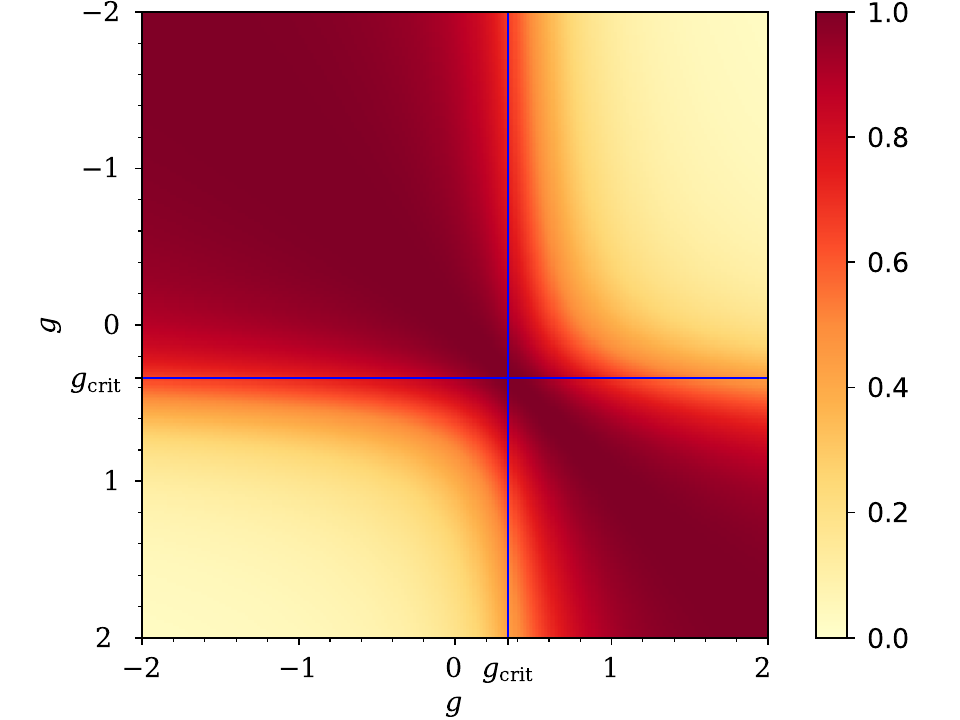}
    \caption{Overlap between the exact ground states for the pairing Hamiltonian sampled from 200 equidistant coupling values $g\in[-2,2]$ for $\Omega=10$ at half-filling $N_\text{occ}=5$.}
    \label{fig:Normmatrix}
\end{figure}

As shown by the second- and third-order MBPT results, the perturbative expansion is only reliable for moderate couplings and fails for larger values. However, the \ecpt{1} results are in good agreement with the exact solution for a wide range of couplings. The EC framework thus effectively resums particle-hole correlations emerging from leading order perturbative corrections to the ground state. In addition, \ecpt{1} and \pCI{2} give very similar results for all coupling values. This is to be expected since both frameworks probe up to two-particle-two-hole correlations in the configuration space, either by a large-scale expansion in Slater determinants in \pCI{2} or by a small-scale expansion of multi-reference states arising from the perturbative state amplitudes in \ecpt{1}.
Here we emphasize that in this work the evaluation of the EC kernel is performed based on a diagrammatic approach and is hence only of polynomially scaling complexity. This makes it possible to compute \ecpt{1} corrections for arbitrary system sizes. In contrast, in Ref.~\cite{Demol2020EC} the perturbative corrections at different orders were computed recursively using an exponentially large configuration basis, which restricts this approach to limited basis spaces $\Omega \leq 20$.

As shown in Fig.~\ref{fig:mbbenchmark}, the \ecpt{1} results perform at a similar level as state-of-the-art nonperturbative coupled-cluster doubles (CCD)~\cite{Hage14RPP,Lie17CompNucPhys} and in-medium similarity renormalization group (IMSRG)~\cite{Herg16PR,Hergert2017} calculations when restricted to the normal phase, which is relevant for ab initio calculations of medium-mass nuclei.
Still the CCD and IMSRG results are consistently closer to the exact solution which we attribute to a more elaborate resummation of higher-order effects that are absent from the simpler \ecpt{1} estimate. Once further corrections are included, \eg{}, using \ecpt{$n\geq 2$}, the variational character of the EC ensures improvement of our results towards the Richardson solution.

The failure of conventional basis-expansion approaches can be understood in terms of the norm matrix shown in Fig.~\ref{fig:Normmatrix} obtained by sampling $n=200$ couplings from the interval $g\in[-2,2]$ and evaluating the overlap of the resulting ground-state eigenvectors.
Matrix elements $|N_{pq}| \approx 1$ indicate similar structures in the ground states, $|\Psi(g_p)\ra$ and $|\Psi(g_q)\ra$.
In the case where $g_p$ and $g_q$ belong to the same regime (either normal or superfluid) the eigenvectors show strong linear dependencies, thus, indicating similar correlations for the two couplings.
However, for $g_p < \gcrit < g_q$ the overlap is significantly reduced yielding almost orthogonal eigenvectors from the two regimes.
As a consequence, one cannot expect a Slater-determinant-based correlation expansion to give accurate results in the superfluid regime~\cite{Lacroix2012}.
This also holds for non-perturbative expansions since the breakdown is related to the use of an improper reference state and not a lack of non-perturbative ladder-type resummations of particle-hole correlations. Figure~\ref{fig:Normmatrix} also nicely illustrates the weak dependence of the eigenstates on the parameters in each regime, which is fundamental to the EC.

\section{Eigenvector continuation as an emulation tool}

Next, we use EC to construct a many-body emulator from a selected set of training data.
Since the ground-state dynamics in the normal and superfluid regimes are fundamentally different, we will investigate the performance of the EC approach as a function of the training data that enter the definition of the EC manifold in Eq.~\eqref{eq:ECman}.
Therefore, three different scenarios characterized by different training sets will be investigated:
\begin{enumerate}
\itemsep0.5em 
\item[i)] a normal training set $\Snormal$, where all training points are in the normal regime, 
\item[ii)] a superfluid training set $\Ssuperfluid$, where all training points are in the superfluid regime,
\item[iii)] a mixed set $\Smixed$, where training points from both regimes are present. 
\end{enumerate}
Naturally, this gives rise to three emulator types: \ECnormal{}, \ECsuperfluid{}, and \ECmixed{}, that have been trained using the corresponding regimes.

\begin{figure}[t]
    \centering
    \includegraphics[width=\columnwidth,clip=]{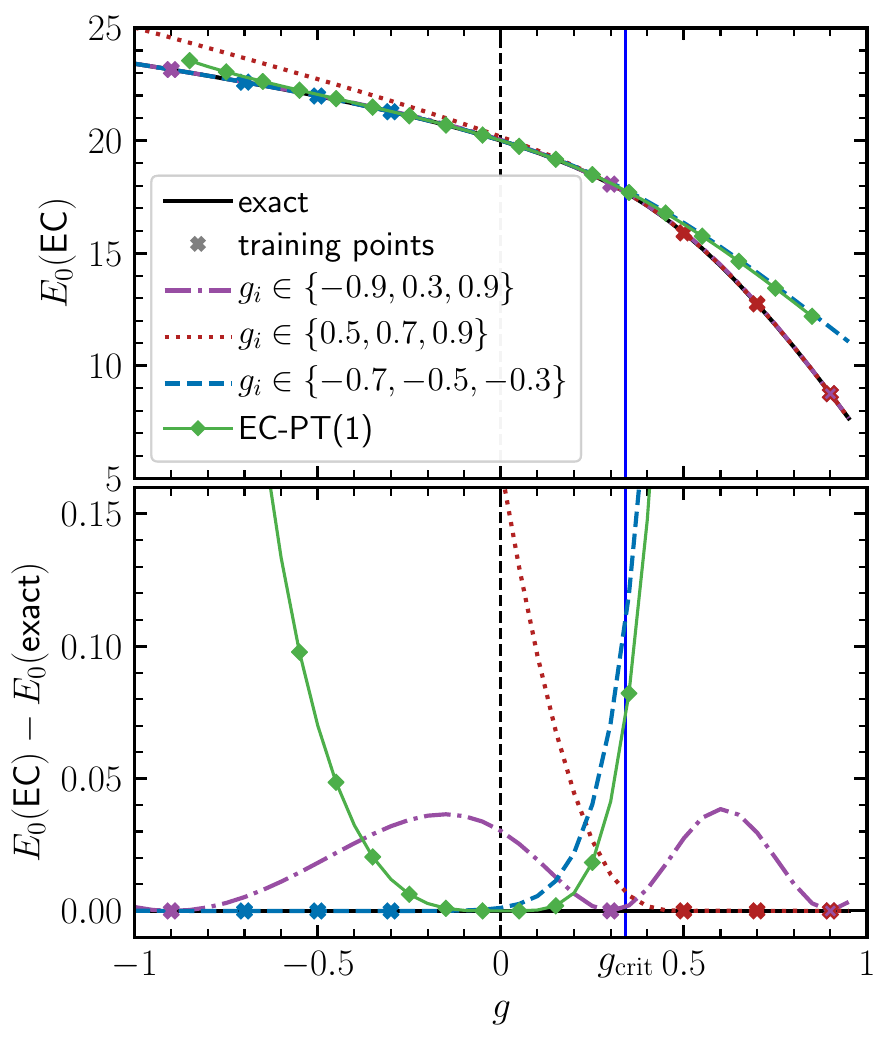}
    \caption{EC results for the ground-state energy (top panel) of the pairing Hamiltonian for three different training sets (dashed, dash-dotted, and dotted lines with three training points indicated in the legend and by the crosses) in comparison to the exact and EC-PT(1) results. The bottom panel shows the deviation from the exact result. Results are given for $\Omega=10$ at half-filling $N_\text{occ}=5$, and $\gcrit=0.34$ is indicated by the vertical line.}
    \label{fig:ECtrain3}
\end{figure}

Figure~\ref{fig:ECtrain3} shows the EC emulator predictions for the ground-state energy for the three different training sets as well as the deviation from the exact result. 
By construction, the error at the training points vanishes because the exact many-body state was used as training vector.
We observe that the \ECnormal{} emulator (dashed line) performs well in the normal regime while incorrectly predicting the superfluid regime. Once the critical coupling is approached the error rapidly increases and no quantitative prediction is possible.
Similarly, \ECsuperfluid{} (dotted line) accurately predicts the superfluid regime while being incapable of capturing the correct trend in the normal regime.
These limitations are overcome by employing a mixed emulator \ECmixed{}: now the normal and superfluid regimes are correctly described and only small deviations appear in between training points.

We conclude that the inclusion of both dynamical particle-hole correlations and static pairing correlations are crucial in the training stage since otherwise the EC manifold probes an insufficient subspace for capturing the Hamiltonian's dynamics at arbitrary coupling values.
These findings are consistent with the form of the norm matrix (see Fig~\ref{fig:Normmatrix}): while in the same regime two vectors have a large overlap, there is nearly no overlap between the normal and the superfluid regime. Therefore, it is not surprising that only selections of training points from both regimes can approximate the solution across both ranges of couplings.
Finally, we compare the EC as an emulator to the \ecpt{1} prediction. Here large errors have to be expected, since only $2$p$2$h-excitations span the \ecpt{1} manifold, lacking important higher-body excitations in the normal phase.
The error can be systematically improved by including higher-order state corrections in \ecpt{n}.
However, the breakdown beyond the critical coupling is associated to the insufficient HF reference state in the presence of superfluid pairing correlations.

Finally, we comment on the impact of enlarging the training set. While for two training points it was not possible to approximate the solution qualitatively over the considered coupling range, three training points were sufficient to approximate the solution. Once a mixed training set is chosen, a further increase in the number of training vectors does not significantly increase the quality of the emulator. This is consistent with the observation from the norm kernel, which reveals strong linear dependencies among vectors from the same regime. Therefore, adding (almost) co-linear vectors will not further improve the emulated results.
The location of training points seems to be more important than the total number of training points. Once the various physical regimes are covered in the training stage, the emulator quantitatively predicts the exact solution very well.
Our general findings agree with the results from Ref.~\cite{Baran2022rbm} that employ an iterative update of the RBM manifold.

\section{Truncation of the training vectors}

\begin{figure}[t]
    \centering
    \includegraphics[width=\columnwidth,clip=]{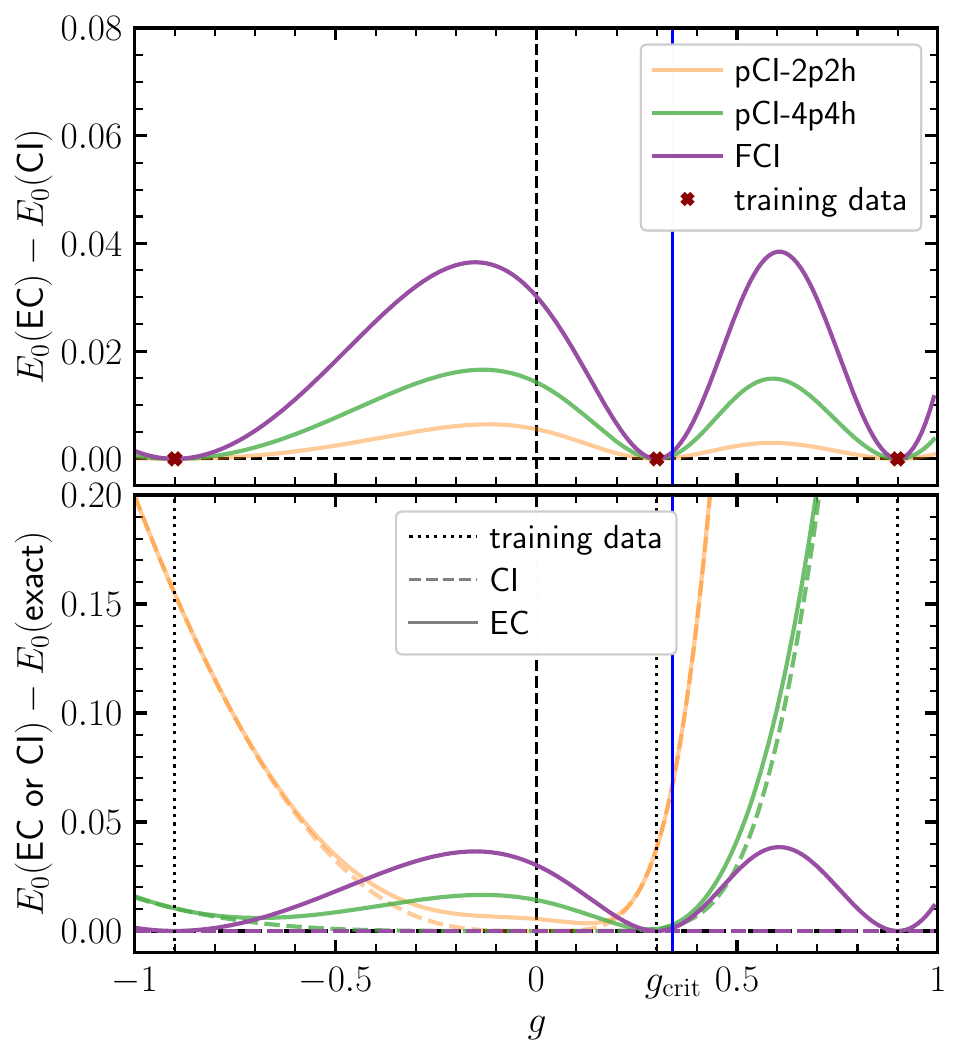}
    \caption{Difference between the EC and pCI-$n$p$n$h ground-state energies (dashed lines in the lower panel) for different truncations (top panel) of the pairing Hamiltonian as well as their deviation from the exact result (bottom panel). The EC is based on the case of the mixed training points $g_i\in\{-0.9,0.3,0.9\}$. Results are given for $\Omega=10$ at half-filling $N_\text{occ}=5$.}
    \label{fig:ECandCIerror}
\end{figure}

The availability of the exact many-body solution for a given parameter set is almost never fulfilled in realistic applications and hence the design of emulators naturally depends on the many-body approximations.
Therefore, the performance of EC emulators has to be validated when approximate training vectors are employed.
To this end, we employ the $n$-particle-$n$-hole pair configuration interaction (pCI-$n$p$n$h) to obtain training data at the 2p2h and 4p4h truncation.
In general, the quality of truncated EC emulators is limited by the quality of the many-body approximations used for the training vectors. Therefore, improvements of the truncated EC results against the exact many-body solution need to be considered with care.

\begin{figure}[t]
    \centering
    \includegraphics[width=\columnwidth,clip=]{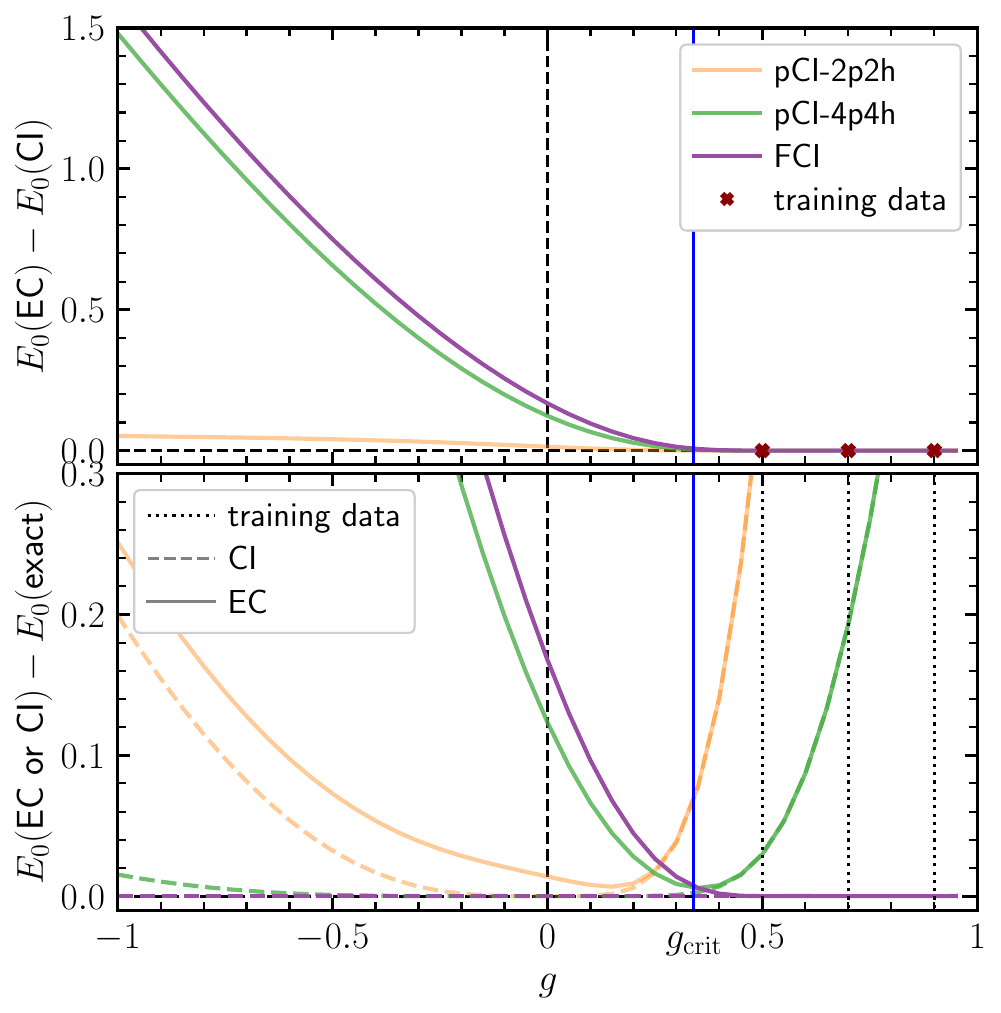}
    \caption{Same as Fig.~\ref{fig:ECandCIerror} but for the superfluid training points $g_i\in\{0.5,0.7,0.9\}$.}
    \label{fig:ECandCIerror2}
\end{figure}

In Fig.~\ref{fig:ECandCIerror} we show the EC ground-state energies for 2p2h-truncated and 4p4h-truncated training data and for the exact (10p10h or FCI) training data for the case of the mixed training set. The training-point truncations in between are not shown, as they differ little from the FCI case. In the top panel of Fig.~\ref{fig:ECandCIerror}, we observe that the difference between the EC and pCI-$n$p$n$h energies increases as the quality of the many-body approximation improves. Comparing truncated EC and pCI-$n$p$n$h results against the exact results shows at small couplings a probably accidental improvement at lower particle-hole truncations, but overall the better the many-body approximation the better is the truncated EC and pCI-$n$p$n$h. This is especially pronounced in the superfluid regime.
To explore this further, we consider in Fig.~\ref{fig:ECandCIerror2} the superfluid training set.
Also for this training set, the difference between the EC and pCI-$n$p$n$h energies is smallest at the 2p2h level. Moreover, for this case of the superfluid training set, the EC at the 4p4h and FCI level does not reproduce the CI (or the exact) solution well when extrapolated to the normal regime.

As can be seen both in the lower panels of Fig.~\ref{fig:ECandCIerror} and Fig.~\ref{fig:ECandCIerror2}, EC better emulates lower CI truncations. This is not surprising, since the size for the EC subspace is three for all three training point truncations. However, while for pCI-2p2h the three-dimensional subspace is used to approximate a 26-dimensional subspace, for full CI it tries to emulate the full 252-dimensional space, which is much larger. This makes emulating full CI a more complex problem.
Therefore, while the more accurate full CI approximation gives the better approximation close to the training points, further away from the training points, it is more difficult for EC to emulate the larger model space.

\section{Summary and conclusions}

In this Letter we applied the EC framework to the pairing Hamiltonian.
In the first part we used EC to resum the perturbative expansion based on a HF state.
In the second part, we designed an EC emulator and tested its sensitivity on the location of training data and many-body approximations for the training vectors.
The EC-resummed perturbative results were in qualitative agreement with the exact Richardson solution even though the underlying perturbative corrections indicated a divergent perturbative expansion. The employed \ecpt{1}-truncation gave similar results as the CI results truncated at 2p2h-level. We expect that our \ecpt{1} results can be efficiently improved by designing a quasiparticle extension built on a BCS reference state~\cite{Lacroix2012,Tichai2020review}.

For the emulator design, the proper choice of training data was crucial.
If the training set was entirely located in the normal (superfluid) regime the predictions for the superfluid (normal) regime gave rise to large errors.
By incorporating training points from both regimes (even just one from the respective other regime) a quantitative prediction of the phase transition was achieved and only moderate errors were encountered. While these errors can be systematically reduced by including training points close to the critical value, the specific uncertainties will generally depend on details like level spacing and system size.
In addition, we explore the CI truncation of the training vectors. This showed that, for a good choice of training points, the higher the truncation of the training vectors the better the EC approximation. Surprisingly, further away from the training data, there is a tendency that lower truncated training points give better results. This could be traced to the EC working better for lower truncated CI in these cases.

Future studies will focus on the power of EC emulators for performing large-scale sensitivity studies with respect to variations of the low-energy couplings in chiral Hamiltonians.
While pioneering work has already been performed~\cite{Ekstroem2019,Jiang:2022oba}, more work is needed to develop emulators for other {\it ab initio} methods and to explore the quality of the emulator in the presence of many-body approximations.
Ultimately, EC provides a powerful framework for global statistical analyses that otherwise remain intractable when exhausting high-dimensional parameter spaces through explicit many-body calculations from first principles.

\section*{Acknowledgements}

A.T.~thanks P. Claeys for useful comments on the numerical solution of the pairing Hamiltonian. This work was supported by the European Research Council (ERC) under the European Union's Horizon 2020 research and innovation programme (Grant Agreement No.~101020842).

\bibliographystyle{apsrev4-1}

\bibliography{strongint}

\end{document}